\documentclass[twocolumn,showpacs,preprintnumbers,prl]{revtex4}
\usepackage{physics}
\usepackage{graphicx}
\usepackage{dcolumn}
\usepackage{bm}
\usepackage{color}

\begin{document}

\title{Revival of Quantum Interference by Modulating the Biphotons}

\author{Chih-Hsiang Wu}
\email{seanasdf49@gmail.com}
\author{Chiao-Kai Liu}
\author{Yi-Cheng Chen}
\author{Chih-Sung Chuu}
\email{cschuu@phys.nthu.edu.tw}
\affiliation{Department of Physics, National Tsing Hua University, Hsinchu 30013, Taiwan\\
Center for Quantum Technology, Hsinchu 30013, Taiwan}

\begin{abstract}  
The possibility to manipulate the wavepackets of single photons or biphotons has enriched quantum optics and quantum information science, with examples ranging from faithful quantum-state mapping and high-efficiency quantum memory to the purification of single photons. Here we demonstrate another fascinating use of wavepacket manipulation on restoring quantum interference. By modulating the photons' temporal wavepacket, we observe the revival of post-selected entanglement that would otherwise be degraded or lost due to poor quantum interference. Our study shows that the amount of the restored entanglement is only limited by the forms of modulation and can achieve full recovery if the modulation function is properly designed. Our work has potential applications in long-distance quantum communication and linear optical quantum computation, particularly for quantum repeaters and large cluster states.
\end{abstract}


\maketitle

\textit{Introduction.} Quantum optics and quantum information science are enriched by the possibility to manipulate the wavepackets of single photons or biphotons. For example, by controlling the waveform of a quantum wavepacket~\cite{Kolchin08}, faithful quantum-state mapping between quantum nodes is conceivable~\cite{Cirac97}, the storage efficiency of single photons in atomic ensembles can approach unity~\cite{Gorshkov07,Zhang12}, and the purity of single photons can be increased~\cite{Feng17}. Likewise, if the phases across the quantum wavepacket are manipulated, one can hide single photons in a noisy environment~\cite{Belthangady10}; photon pairs can also behave like fermions~\cite{Specht09}. With these fascinating opportunities present, it is noteworthy that many of them can be realized with temporally long single photons or biphotons from atomic ensembles~\cite{Balic05,Du08}, resonant spontaneous parametric down-conversion~\cite{Kuklewicz06,Bao08,Scholz09,Wolfgramm11,Wu17}, or cavity quantum electrodynamics~\cite{Kuhn02,Keller04,McKeever04,Thompson06}.

In this Letter we demonstrate another fascinating use of wavepacket manipulation: by modulating the biphotons' wavepacket, we restore the quantum interference and post-selected entanglement that would otherwise be destroyed by the photons' distinguishability. Quantum interference and entanglement, apart from the fundamental interest, are at the heart of photonic quantum technologies with examples ranging from quantum communication \cite{Ekert91,Bennet92,Bennett93,Briegel98,Lo12} and quantum computation~\cite{Knill01} to quantum random number generation \cite{Pironio10,Liu18}. In particular, the storage of entanglement in quantum memories plays a critical role in long-distance quantum communication to implement the quantum repeater \cite{Briegel98} and in linear optical quantum computation to generate large cluster states~\cite{Browne05}, where temporally long biphotons with subnatural linewidth are favorable for achieving the optimal performance. To generate entanglement, two-photon interference \cite{Ou88,Shih88} has been the workhorse in many quantum optics experiments and applications~\cite{Kok07,Sangouard11,Pan12}. By sending non-polarization-entangled biphotons (for example, from parametric down-conversion) onto a beam splitter, one may entangle the photons exiting through different ports in the polarization degree of freedom. However, if the frequencies of the biphotons are dissimilar, the interference visibility diminishes; the entanglement then declines or even disappears. Here we report the unexpected finding of restoring the lost two-photon interference as well as the resulting entanglement and nonlocality by modulating the biphotons' temporal wavepacket. Our study shows that the amount of restored entanglement is only limited by the forms of modulation and can achieve full recovery if the modulation function is properly designed.

\begin{figure}[t]
\centering
\includegraphics[width=0.92 \linewidth]{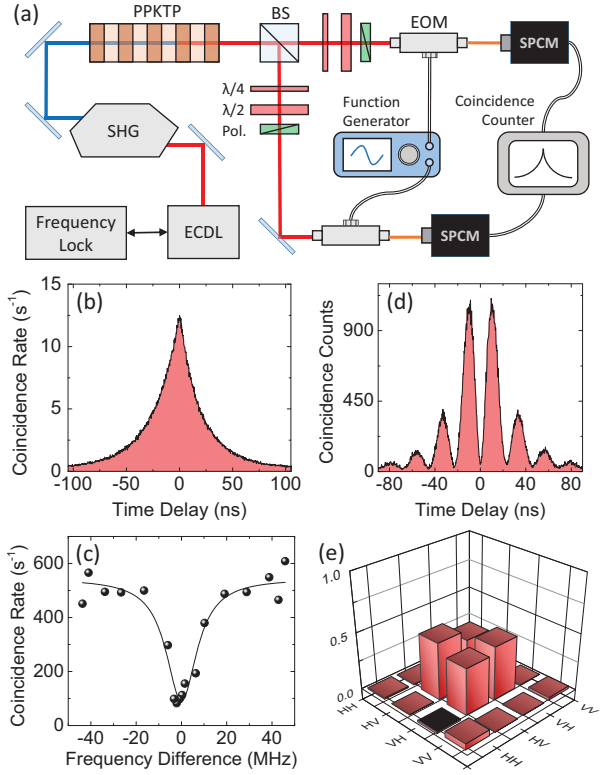}
\caption{\label{fig1} (color online) (a) Schematic of the experimental setup. ECDL: external cavity diode laser, SHG: second harmonic generation, BS: beamsplitter, Pol.: polarizers, $\lambda /2$: half-wave plates, $\lambda /4$: quarter-wave plates, EOM: electro-optic modulators, SPCM: single-photon counting modules. (b) The temporal wavepacket of the biphotons is a double exponential with exponential time constants of 21 and 24 ns. (c) Hong-Ou-Mandel interference shows the distinguishability increasing with the frequency difference of the biphotons. (d) Time-resolved two-photon interference is exploited to measure the frequency difference of the biphoton. In this figure, the frequencies of the photon pair are differed by 43 MHz, resulting in a beat at the same frequency. (e) The tomographic reconstruction of the density matrix for polarization-entangled photons at degeneracy.}
\end{figure}

\textit{Entangled photons with controllable waveforms.} Fig.~\ref{fig1}(a) illustrates the experimental setup for demonstrating the revival of quantum interference and entanglement. Biphotons are first generated by doubly resonant parametric down-conversion in a monolithic KTP crystal \cite{Wu17,Chuu12,supplementary}, with a temporally long wavepacket [Fig.~\ref{fig1}(b)] allowing the arbitrary shaping of the waveforms. The orthogonally polarized photon pairs are then entangled in the polarization degree of freedom by selecting pairs exiting through different ports of the beamsplitter. The frequency difference of the biphotons, which can be tuned via the phase-matching condition by adjusting the pump frequency and crystal temperature, determines the distinguishability in two-photon interference as well as the degree of entanglement. This is manifested in Fig.~\ref{fig1}(c), where the coincidence counts in the Hong-Ou-Mandel interferometry~\cite{supplementary,Hong87}--an indication of the distinguishability--increases with the frequency difference. When the biphotons are degenerate, the indistinguishability in the two-photon interference through the beamsplitter results in the polarization-entangled state $(\ket{H}_a\ket{V}_b+\ket{V}_a\ket{H}_b)/\sqrt{2}$ with \textit{a} and \textit{b} denoting the ports each photon exits. To precisely control the frequency difference, time-resolved two-photon interference as in Fig.~\ref{fig1}(d) is carried out to measure the beat frequency \cite{supplementary}. In the case of degenerate biphotons, the beat vanishes and the real part of the density matrix $\rho$ reconstructed by the quantum state tomography is shown in Fig.~\ref{fig1}(e).

\begin{figure}[t]
\centering
\includegraphics[width=0.9 \linewidth]{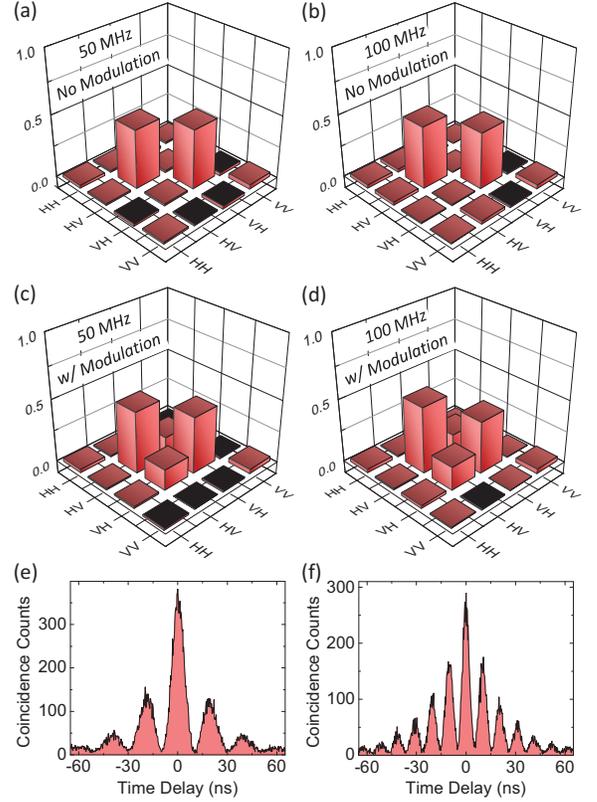}
\caption{\label{fig2} (color online) The density matrices at frequency differences of (a) 50 MHz and (b) 100 MHz show the loss of coherence and entanglement. The density matrices at frequency differences of (c) 50 MHz and (d) 100 MHz show the recovery of coherence and entanglement after shaping the wavepackets as in (e) and (f), respectively.}
\end{figure}

\begin{figure}[t]
\centering
\includegraphics[width=0.71 \linewidth]{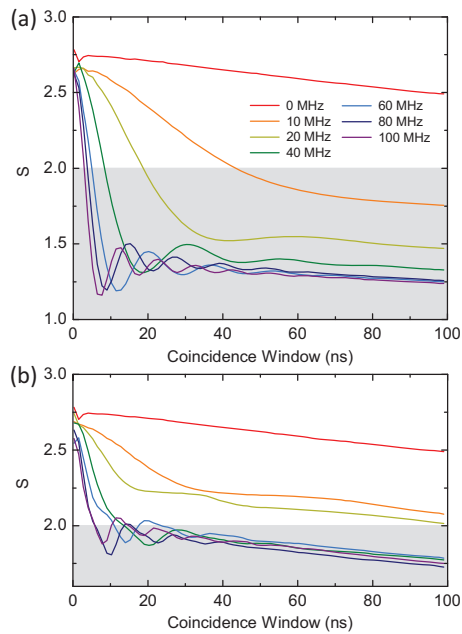}
\caption{\label{fig3} (color online) (a) The CHSH inequality $|S| \leq 2$ requires shorter coincidence windows to be violated when the frequency difference of the biphoton increases. (b) If the temporal wavepacket is modulated, the inequality with nondegenerate biphotons can be violated at larger coincidence windows.}
\end{figure}

\textit{Characterization of entanglement.} To characterize the entanglement, we exploit the concurrence \cite{Wootters98} as the measure. More specifically, we obtain the square root of the eigenvalues of $\rho(\sigma_y \otimes \sigma_y)\rho^*(\sigma_y \otimes \sigma_y)$ in descending order, $\left\{ \sqrt{\lambda_1},\sqrt{\lambda_2},\sqrt{\lambda_3},\sqrt{\lambda_4} \right\}$. The concurrence is then calculated by $C={\rm max}(0,\sqrt{\lambda_1}-\sqrt{\lambda_2}-\sqrt{\lambda_3}-\sqrt{\lambda_4})$ with values ranging from 0 for non-entangled states to 1 for maximally entangled states. Using this entanglement measure, the density matrix of degenerate biphotons in Fig.~\ref{fig1}(e) gives $C=0.71$ with a purity ${\rm \textit{Tr}}[\rho^2]$ = 0.81, thus verifying the presence of polarization entanglement. Figs.~\ref{fig2}(a) and (b) show the tomographic reconstruction of the density matrices for nondegenerate biphotons with a frequency difference of 50 and 100 MHz, respectively. For both frequency differences, the concurrence and the purity are 0 and 0.45, respectively. Thus, as the frequency difference increases, the concurrence and the degree of entanglement declines rapidly. Similar decline of entanglement can also be observed in the violation of the Clauser-Horne-Shimony-Holt (CHSH) inequality~\cite{Clauser69}, $|S| \leq 2$, which places constraints on the value \textit{S} of a combination of four polarization correlation probabilities analyzed by two possible settings for each photon. If the CHSH inequality is violated, quantum entanglement is necessary to explain the correlations or nonlocality. Fig.~\ref{fig3}(a) shows the measured \textit{S} \cite{supplementary} at frequency differences of 0, 10, 20, 40, 60, 80, and 100~MHz with the coincidence window varied from 0 to 100 ns. At degeneracy (red curve), the CHSH inequality is violated and nonlocality is observed no matter how large the coincidence window is. However, when the frequency difference increases, it requires a smaller coincidence window and thus higher indistinguishability in the two-photon interference to violate the CHSH inequality and observe the nonlocality.

\textit{Revival of entanglement and nonlocality.} Surprisingly, the lost entanglement or nonlocality can be restored by shaping the biphotons. To demonstrate this aspect, we modulate each photon of the pair synchronously with electro-optic modulator at half of the frequency difference of the pair so that the biphoton wavepacket is modulated by the convolution of these modulation functions at the frequency of pair's frequency difference. Fig.~\ref{fig2}(e) shows the wavepacket of biphotons, which has a frequency difference of 50 MHz and is modulated with a triangular function at the same frequency. By tomographically reconstructing the density matrix in Fig.~\ref{fig2}(c), we observe the increase in both the concurrence $C=0.28$ and purity ${\rm \textit{Tr}}[\rho^2]$ = 0.5, thus manifesting the revival of entanglement. 

The biphoton modulation, as implied by the nonclassical time correlation of the biphotons, is equivalent to modulating one photon of the pair conditionally (on the detection of another photon) by the convolution of modulation functions. It is thus possible to observe the revival of entanglement by modulating one photon of the pair. To demonstrate this nonclassical effect, we prepare biphotons with a frequency difference of 100 MHz and modulate the signal photons conditionally \cite{supplementary} with a cosinusoidal function at 100 MHz [Fig.~\ref{fig2}(f)]. The density matrix in such an instance is shown in Fig.~\ref{fig2}(d). The concurrence $C=0.32$ and purity ${\rm \textit{Tr}}[\rho^2]$ = 0.5 both increase compared to those of the unmodulated biphotons. In Fig.~\ref{fig3}(b) we also examine the violation of the CHSH inequality when the wavepacket is modulated. Compared to the unmodulated cases in Fig.~\ref{fig3}(a), the inequality with nondegenerate biphotons can now be violated at larger coincidence windows. For example, the inequality with a frequency difference of 20 MHz (yellow curve), which would not be violated previously for coincidence windows larger than 19 ns, is violated all the way up to 100 ns. The nonlocality of the entangled photons is thus restored by shaping the wavepacket.

\begin{figure}[t]
\centering
\includegraphics[width=0.93 \linewidth]{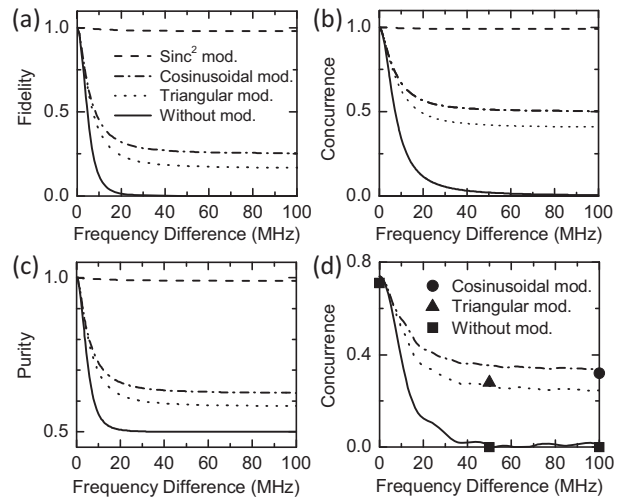}
\caption{\label{fig4} (a) The fidelity of the two-photon wavefunctions contributing to quantum interference is reduced by the frequency difference (solid curve) but can be recovered by modulating the biphotons. (b) The entanglement, quantified here by concurrence, is lost at large frequency differences (solid curve). However, if the biphoton or single photon is modulated, the entanglement remains at large frequency difference. (c) The purity of the entangled state also increases. (d) The measured concurrence (solid squares, triangle and circle) is in good agreement with the calculated concurrence (curves) by taking into account the asymmetry in wavepacket, accidental coincidence, the nonideal split ratio of the beamsplitter and the imperfect preparation of measurement bases.}
\end{figure}

\textit{Why it works?} The revival of quantum interference, entanglement, or nonlocality by shaping the photons' wavepacket can be understood in the following ways. From the time-domain point of view, the modulation at the frequency of biphoton's frequency difference ensures a constant phase between the two-photon states in two-photon interference. From the frequency-domain point of view, the modulation at the frequency of biphoton's frequency difference generates signal (or idler) photons at the carrier frequency or sidebands of the idler (or signal) photons. In both viewpoints, the indistinguishability in two-photon interference increases as a consequence of modulation, thus improving the quality of entanglement and the ability to observe nonlocality. This can be seen in Fig.~\ref{fig4}(a), where we calculate the fidelity of two-photon wavefunctions contributing to the quantum interference. Without the modulation, the fidelity (and therefore the interference visibility) decreases with the frequency difference. If the biphotons are modulated by properly designed functions (details of various functions are given below), the fidelity is recovered for all frequency differences.

\begin{figure}[t]
\centering
\includegraphics[width=0.71 \linewidth]{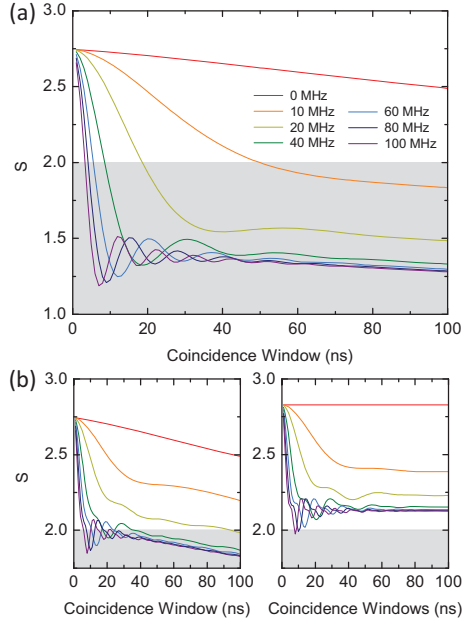}
\caption{\label{fig5} Compared to the CHSH inequality with unmodulated wavepacket in (a), the inequality with modulated wavepacket in (b) (left panel) can be violated with a longer coincidence window at larger frequency difference. The calculation considers the imperfection in experiment, including the accidental coincidence. In the absence of experimental imperfection, the violation in (b) (right panel) is independent of the size of coincidence window when the wavepacket is modulated.}
\end{figure}

More rigorously, we consider the following quantum state, 
\begin{eqnarray}
\label{eq:initial_state}
\ket{\psi} &=& \frac{1}{\sqrt{2}} \int \int dt d\tau [ \phi_{HV}(\tau) \hat{a}^{\dagger}_{1H}(t)\hat{a}^{\dagger}_{2V}(t+\tau) + \\ \nonumber
&& \ e^{i \Delta \omega \tau} \phi_{VH}(\tau) \hat{a}^{\dagger}_{1V}(t) \hat{a}^{\dagger}_{2H}(t+\tau)] e^{i[\omega_s t+\omega_i (t+\tau)]} \ket{0},
\end{eqnarray}
where $\phi_{HV}(\tau)$ and $\phi_{VH}(\tau)$ are the two-photon wavefunctions, $\hat{a}^{\dagger}_{ij}(t)=m_i(t)\hat{a}^{\dagger}_{ij,0}(t)$ ($\hat{a}^{\dagger}_{ij,0}(t)$) is the creation operator of the \textit{j}-polarized photon at \textit{i}th port after (before) the amplitude modulation $m_i(t)$,  and $\Delta \omega /2\pi = (\omega_s-\omega_i)/2\pi$ is the frequency difference of the biphoton. In our experiment, $\phi_{HV}(\tau) = \phi_{VH}(-\tau) = e^{-|\tau|/2\tau_0}$ for the nondegenerate biphotons exiting through different ports of the beam splitter. Taking $m_i(t)=1$ (no modulation), we see that the coherence $\zeta(\Delta \omega) = (1/4\tau_0) \int_{-\infty}^{\infty} e^{-|\tau|/\tau_0} e^{-i\Delta \omega \tau} d\tau = 1/2(1+\theta^2)$ with $\theta = \Delta \omega \tau_0$ in the density matrix,
\begin{eqnarray}
&&\rho = \left( 
\begin{array}{cccc}
 0 & 0 & 0 & 0 \\
 0 & 1/2 & \zeta(\Delta \omega) & 0 \\
 0 & \zeta(\Delta \omega) & 1/2 & 0 \\
 0 & 0 & 0 & 0
\end{array} \right),
\label{eq:rho_initial}
\end{eqnarray}
is gradually lost as the frequency distinguishability increases. When $\Delta \omega \gg \tau_0^{-1}$, the coherence as well as the concurrence $C=2 \zeta(\Delta \omega)$ [solid curve in Fig.~\ref{fig4}(b)] vanishes; the entanglement does not exist anymore. However, if we modulate the signal and idler photons by synchronized square-wave functions $|m_i(t)|^2=|S(t)|^2$ (the biphoton wavepacket is accordingly modulated by triangular function $M(\tau)=(1/T)\int_0^T |S(t)|^2 |S(t+\tau)|^2 dt $ at twice the frequency with the modulation period denoted by $T$), the coherence $\zeta(\Delta \omega) = \int_{-\infty}^{\infty} M(\tau) e^{-|\tau|/\tau_0} e^{-i\Delta \omega \tau} d\tau / [2 \int_{-\infty}^{\infty} M(\tau) e^{-|\tau|/\tau_0} d\tau] = (\pi x - \theta + \pi x \theta^2 + \theta^3)/[2x(\pi - x \theta) (1+\theta^2)^2]$ with $x = {\rm tanh}(\pi/2 \theta)$, which peaks at the modulation frequency of $\Delta \omega /2 \pi$, remains finite even if $\Delta \omega \gg \tau_0^{-1}$. The concurrence [dot curve in Fig.~\ref{fig4}(b)] and thus the entanglement are restored at large frequency differences. The corresponding change of the purity is also shown in Fig.~\ref{fig4}(c). In the case where one photon is conditionally modulated (or the biphoton wavepacket is modulated) by a cosinusoidal function $(1+e^{i \Delta \omega \tau})/2$, both the coherence $\zeta(\Delta \omega) = (2+7\theta^2+2\theta^4)/2(2+\theta^2)(1+4\theta^2)$ and concurrence [dash-dot curve in Fig.~\ref{fig4}(b)] survive at large frequency difference as well; the entanglement is restored again. We note that the concurrence at large frequency differences of $\sim$100 MHz does not place a limitation on the entanglement restored by shaping the biphotons. The amount of entanglement that can be restored depends on the modulation function used. For example, if the biphotons are modulated by periodic sinc$^2$ function $\int_{-\infty}^{\infty} |m(t)|^2 |m(t,\tau)|^2 dt = |(1/s)\sum_{n=1}^s{\rm exp}(i n \Delta \omega \tau)|^2$ with $s=100$ (which reduces to the cosinusoidal function if $s=2$), the entanglement can be restored nearly to its quality at degeneracy [dash curve in Fig.~\ref{fig4}(b)].

In Fig.~\ref{fig4}(d) we calculate the concurrence before (solid curve) and after the photons are shaped by the triangular (dot curve) and cosinusoidal (dash-dot curve) functions while taking into account the asymmetry of the double exponential waveform in the biphoton wavepacket, the accidental coincidence, the beamsplitter with split ratio deviated from 50:50, and the imperfect preparation of measurement bases in our experiments. Compared to the experimental data (solid squares, triangle, and circle), the calculated concurrences are all in good agreement. In Figs.~\ref{fig5}(a) and \ref{fig5}(b) (left panel) we calculate the $S$-value to examine the violation CHSH inequality before and after modulating the wavepacket, respectively, by considering the imperfection in experiment. They are also in good agreement with the measurements in Fig.~\ref{fig3}. We note that the decrease of $S$ at large coincidence windows is mainly due to the accidental coincidence. As shown in Fig.~\ref{fig5}(b) (right panel), where the experimental imperfection is assumed to be absent, the observation of nonlocality is independent of the size of coincidence window when the wavepacket is modulated.

\textit{Conclusion.} In summary, by modulating the biphoton's temporal wavepacket, we have observed the revival of quantum interference, entanglement and nonlocality that would otherwise be degraded or lost due to the frequency distinguishability of biphotons. Our study also shows that the amount of restored entanglement is only limited by the forms of modulation and can achieve full recovery if the modulation function is properly designed. Quantum entanglement is at the heart of many photonic quantum technologies \cite{Ekert91,Bennet92,Bennett93,Knill01}. The storage of entanglement in quantum memories, which requires narrowband biphotons, is particularly important for realizing quantum repeaters \cite{Briegel98} or large cluster states \cite{Browne05}. Our work thus has potential applications for long-distance quantum communication and linear optical quantum computation. We note that the method present here may also be applied to restore the two-photon interference between independent nondegenerate photons, which can be seen by replacing the biphoton wavefunction in Eq.~(1) with the product of single-photon wavefunctions. The distinction between restoring the quantum interferences of single photons and biphotons is that the former necessitates well-defined time origin for the modulation while the latter does not. 

The authors thank S. E. Harris for helpful discussions. This work was supported by the Ministry of Science and Technology, Taiwan (107-2112-M-007-004-MY3, 107-2627-E-008-001 and 107-2745-M-007-001).


\begin{thebibliography}{99}

\bibitem{Kolchin08} P. Kolchin, C. Belthangady, S. Du, G. Y. Yin, and S. E. Harris, Phys. Rev. Lett. \textbf{101}, 103601 (2008).

\bibitem{Cirac97} J. I. Cirac, P. Zoller, H. J. Kimble, and H. Mabuchi, Phys. Rev. Lett. \textbf{78}, 3221 (1997).

\bibitem{Gorshkov07} A. V. Gorshkov, A. Andr$\acute{\rm e}$, M. Fleischhauer, A. S. S{\o}rensen, and M. D. Lukin, Phys. Rev. Lett. \textbf{98}, 123601 (2007).

\bibitem{Zhang12} S. Zhang, C. Liu, S. Zhou, C.-S. Chuu, M. M. T. Loy, and S. Du, Phys. Rev. Lett. \textbf{109}, 263601 (2012).

\bibitem{Feng17} S.-W. Feng, C.-Y. Cheng, C.-Y. Wei, J.-H. Yang, Y.-R. Chen, Y.-W. Chuang, Y.-H. Fan, and C.-S. Chuu, Phys. Rev. Lett. \textbf{119}, 143601 (2017).

\bibitem{Belthangady10} C. Belthangady, C.-S. Chuu, I. A. Yu, G. Y. Yin, J. M. Kahn, and S. E. Harris, Phys. Rev. Lett. \textbf{104}, 223601 (2010).

\bibitem{Specht09} H. P. Specht, J. Bochmann, M. Mücke, B. Weber, E. Figueroa, D. L. Moehring, and G. Rempe, Nat. Photonics \textbf{3}, 469 (2009).

\bibitem{Balic05} V. Bali${\rm \acute{c}}$, D. A. Braje, P. Kolchin, G. Y. Yin, and S. E. Harris, Phys. Rev. Lett. \textbf{94}, 183601 (2005).

\bibitem{Du08} S. Du, P. Kolchin, C. Belthangady, G. Y. Yin, and S. E. Harris, Phys. Rev. Lett. \textbf{100}, 183603 (2008).

\bibitem{Kuklewicz06} C. E. Kuklewicz, F. N. C. Wong, and J. H. Shapiro, Phys. Rev. Lett. \textbf{97}, 223601 (2006).

\bibitem{Bao08} X.-H. Bao, Y. Qian, J. Yang, H. Zhang, Z.-B. Chen, T. Yang, and J.-W. Pan, Phys. Rev. Lett. \textbf{101}, 190501 (2008).

\bibitem{Scholz09} M. Scholz, L. Koch, and O. Benson, Phys. Rev. Lett. \textbf{102}, 063603 (2009).

\bibitem{Wolfgramm11} F. Wolfgramm, Y. A. de Icaza Astiz, F. A. Beduini, A. Cer$\grave{\rm e}$, and M. W. Mitchell, Phys. Rev. Lett. \textbf{106}, 053602 (2011).

\bibitem{Wu17} C.-H. Wu, T.-Y. Wu, Y.-C. Yeh, P.-H. Liu, C.-H. Chang, C.-K. Liu, T. Cheng, and C.-S. Chuu, Phys. Rev. A \textbf{96}, 023811 (2017).

\bibitem{Kuhn02} A. Kuhn, M. Hennrich, and G. Rempe, Phys. Rev. Lett. \textbf{89}, 067901 (2002).

\bibitem{Keller04} H. P. Keller, B. Lange, K. Hayasaka, W. Lange, and H. Walther, Nature (London) \textbf{431}, 1075 (2004).

\bibitem{McKeever04} J. McKeever, A. Boca, A. D. Boozer, R. Miller, J. R. Buck, A. Kuzmich, and H. J. Kimble, Science \textbf{303}, 1992 (2004).

\bibitem{Thompson06} J. K. Thompson, J. Simon, H. Loh, and V. Vuleti${\rm \acute{c}}$, Science \textbf{313}, 74 (2006).

\bibitem{Ekert91} A. K. Ekert, Phys. Rev. Lett. \textbf{67}, 661 (1991).

\bibitem{Bennet92} C. H. Bennett, G. Brassard, and N. D. Mermin, Phys. Rev. Lett. \textbf{68}, 557 (1992).

\bibitem{Bennett93} C. H. Bennett, G. Brassard, C. Cr\'epeau, R. Jozsa, A. Peres, and W. K. Wootters, Phys. Rev. Lett. \textbf{70}, 1895 (1993).

\bibitem{Briegel98} H.-J. Briegel, W. D\"ur, J. I. Cirac, and P. Zoller, Phys. Rev. Lett. \textbf{81}, 5932 (1998).

\bibitem{Lo12} H.-K. Lo, M. Curty, and B. Qi, Phys. Rev. Lett. \textbf{108}, 130503 (2012).

\bibitem{Knill01} E. Knill, R. Laflamme and G. J. Milburn, Nature (London) \textbf{409}, 46 (2001).

\bibitem{Pironio10} S. Pironio, A. Acín, S. Massar, A. Boyer de la Giroday, D. N. Matsukevich, P. Maunz, S. Olmschenk, D. Hayes, L. Luo, T. A. Manning, and C. Monroe, Nature (London) \textbf{464}, 1021 (2010).

\bibitem{Liu18} Y. Liu, Q. Zhao, M.-H. Li, J.-Y. Guan, Y. Zhang, B. Bai, W. Zhang, W.-Z. Liu, C. Wu, X. Yuan, H. Li, W. J. Munro, Z. Wang, L. You, J. Zhang, X. Ma, J. Fan, Q. Zhang, and J.-W. Pan, Nature (London) \textbf{562}, 548 (2018).

\bibitem{Browne05} D. E. Browne and T. Rudolph, Phys. Rev. Lett. \textbf{95}, 010501 (2005).

\bibitem{Ou88} Z. Y. Ou and L. Mandel, Phys. Rev. Lett. \textbf{61}, 50 (1988).

\bibitem{Shih88} Y. H. Shih and C. O. Alley, Phys. Rev. Lett. \textbf{61}, 2921 (1988).






\bibitem{Kok07} P. Kok, W. J. Munro, K. Nemoto, T. C. Ralph, J. P. Dowling, and G. J. Milburn, Rev. Mod. Phys. \textbf{79}, 135 (2007).

\bibitem{Sangouard11} N. Sangouard, C. Simon, H. de Riedmatten, and N. Gisin, Rev. Mod. Phys. \textbf{83}, 33 (2011).

\bibitem{Pan12} J.-W. Pan, Z.-B. Chen, C.-Y. Lu, H. Weinfurter, A. Zeilinger, and M. \.Zukowski, Rev. Mod. Phys. \textbf{84}, 777 (2012).

\bibitem{supplementary} See Supplemental Material for more details of the generation and measurements of entangled photons, which includes additional Refs. [34-46].

\bibitem{Chuu11} C.-S. Chuu and S. E. Harris, Phys. Rev. A \textbf{83}, 061803(R) (2011).

\bibitem{Guo17} X. Guo, Y. Mei, and S. Du, Optica \textbf{4}, 388 (2017).

\bibitem{Kwiat92} P. G. Kwiat, A. M. Steinberg, and R. Y. Chiao, Phys. Rev. A \textbf{45}, 7729 (1992).

\bibitem{Kuklewicz04} C. E. Kuklewicz, M. Fiorentino, G. Messin, F. N. C. Wong, and J. H. Shapiro, Phys. Rev. A \textbf{69}, 013807 (2004).

\bibitem{Zhang11} H. Zhang, X.-M. Jin, J. Yang, H.-N. Dai, S.-J. Yang, T.-M. Zhao, J. Rui, Y. He, X. Jiang, F. Yang, G.-S. Pan, Z.-S. Yuan, Y. Deng, Z.-B. Chen, X.-H. Bao1, S. Chen, B. Zhao, and J.-W. Pan, Nat. Photonics \textbf{5}, 628 (2011).

\bibitem{Dai12} H.-N. Dai, H. Zhang, S.-J. Yang, T.-M. Zhao, J. Rui, Y.-J. Deng, L. Li, N.-L. Liu, S. Chen, X.-H. Bao, X.-M. Jin, B. Zhao, and J.-W. Pan, Phys. Rev. Lett. \textbf{108}, 210501 (2012).

\bibitem{Fink17} M. Fink, A. Rodriguez-Aramendia, J. Handsteiner, A. Ziarkash, F. Steinlechner, T. Scheidl, I. Fuentes, J. Pienaar, T. C. Ralph, and R. Ursin, Nat. Commun. \textbf{8}, 15304 (2017).

\bibitem{Eberhard93} P. H. Eberhard, Phys. Rev. A \textbf{47}, R747 (1993).

\bibitem{DeCaro94} L. De Caro and A. Garuccio, Phys. Rev. A \textbf{50}, R2803 (1994).

\bibitem{Popescu97} S. Popescu, L. Hardy, and M. \.Zukowski, Phys. Rev. A \textbf{56}, R4353 (1997).

\bibitem{Zukowski99} M. \.Zukowski, D. Kaszlikowski, and E. Santos, Phys. Rev. A \textbf{60}, R2614 (1999).

\bibitem{Thyagarajan09} K. Thyagarajan, J. Lugani, S. Ghosh, K. Sinha, A. Martin, D. B. Ostrowsky, O. Alibart, and S. Tanzilli, Phys. Rev. A \textbf{80}, 052321 (2009).

\bibitem{Kim06} T. Kim, M. Fiorentino, and F. N. C. Wong, Phys. Rev. A \textbf{73}, 012316 (2006).

\bibitem{Chuu12} C.-S. Chuu, G. Y. Yin, and S. E. Harris, Appl. Phys. Lett. \textbf{101}, 051108 (2012).

\bibitem{Hong87} C. K. Hong, Z. Y. Ou, and L. Mandel, Phys. Rev. Lett. \textbf{59}, 2044 (1987).


\bibitem{Wootters98} W. K. Wootters, Phys. Rev. Lett. \textbf{80}, 2245 (1998).

\bibitem{Clauser69} J. F. Clauser, M. A. Horne, A. Shimony, and R. A. Holt, Phys. Rev. Lett. \textbf{23}, 880 (1969).







\end{thebibliography}
\end{document}